\documentclass{optica-article}

\journal{opticajournal} 

\articletype{Research Article}

\usepackage{lineno}

\usepackage[normalem]{ulem}
\usepackage{threeparttable}
\usepackage{subcaption}
\begin{document}

\title{Broadband anti-reflection coating for sub-terahertz optics using dielectric multilayers}

\author{Toyo Naganuma,\authormark{1,2}
        Shinsuke Uno,\authormark{3}
        Shuhei Inoue,\authormark{2,4}
        Kazuki Watanabe,\authormark{2,5}
        Tatsuya Takekoshi,\authormark{6}
        Takeshi Sakai,\authormark{1}
        Tai Oshima\authormark{2,5,*}
}

\address{\authormark{1}Graduate School of Informatics and Engineering, The University of Electro-Communications, 1-5-1 Chofugaoka, Chofu, Tokyo 182-8585, Japan\\
\authormark{2}National Astronomical Observatory of Japan, 2-21-1 Osawa, Mitaka, Tokyo 181-8588, Japan\\
\authormark{3}RIKEN Center for Advanced Photonics, Saitama 351-0198, Japan\\
\authormark{4}Institute of Astronomy, Graduate School of Science, The University of Tokyo, 2-21-1 Osawa, Mitaka, Tokyo 181-0015, Japan\\
\authormark{5}Graduate Institute for Advanced Studies (SOKENDAI), 2‑21‑1 Osawa, Mitaka, Tokyo 181‑8588, Japan\\
\authormark{6}Kitami Institute of Technology, 165 Koen‑cho, Kitami, Hokkaido 090‑8507, Japan\\
}

\email{\authormark{*}tai.oshima@nao.ac.jp}

{\noindent \copyright 2026 Optica Publishing Group. One print or electronic copy may be made for personal use only. Systematic reproduction and distribution, duplication of any material in this paper for a fee or for commercial purposes, or modifications of the content of this paper are prohibited.}

{\noindent This article is published in 
Applied Optics Vol. 65, Issue 8, pp. 2468-2474 (2026), and may be found at https://doi.org/10.1364/AO.582598.}

\begin{abstract*}
Sub-terahertz astronomy requires instruments capable of simultaneous observations across multiple spectral bands, motivating the development of 
broadband anti-reflection coatings (ARCs).
We investigated low-loss dielectrics with refractive indices suitable for multilayer ARCs on polyethylene optical elements and identified candidates that partially meet the requirements.
To address the remaining gaps in available refractive indices, we
applied dielectric multilayer synthesis by combining newly identified thin materials with controlled bonding to realize the required effective refractive indices.
As a result, the fabricated 5-layer ARC achieved reflection losses of 0.2\% (average) and 3.2\% (maximum) over 130--710~GHz.
\end{abstract*}

\section{INTRODUCTION}
\label{sec:intro}

Sub-THz astrophysical telescopes probe the structure and chemical evolution of the Universe through measurements of the spectral energy distribution of galaxy clusters and star-forming regions.
Advances in cryogenic detector arrays over the past decade have enabled simultaneous two- to three-color imaging with ground-based radio telescopes, producing high-quality results (e.g., SPT~\cite{Carlstrom2011SPT}, ACT~\cite{Swetz2011ACT}, ASTE~\cite{Takekoshi2012}, SCUBA-2~\cite{Holland2013}, NIKA2~\cite{Adam2018NIKA2}). 
A natural next step is to increase the number of spectral bands to cover more than one octave, which is 
useful
for measuring galaxy cluster motions (via the kinetic Sunyaev--Zel'dovich effect) and determining redshifts of distant galaxies~\cite{takeuchi2001impact,deBernardis2012low}.

We are developing a next-generation sub-THz multicolor camera, GLTCAM~\cite{Inoue2026}, for the 12-m Greenland Telescope (GLT~\cite{2023Chen}), to conduct wide-field surveys of the early universe.  
The camera will simultaneously cover six spectral bands (150, 220, 270, 350, 450, and 670~GHz), spanning 130--710~GHz (138\% fractional bandwidth) over an 18-arcmin field of view~\cite{takekoshi2022material}.  
Here, the fractional bandwidth is defined as $(f_{\max}-f_{\min})/f_{\mathrm{center}}$, where $f_{\min}$ and $f_{\max}$ denote the lower and upper edges of the frequency range, and $f_{\mathrm{center}}$ the center frequency.
The GLTCAM
cryostat requires a large-aperture ($300$-mm diameter) vacuum window that must withstand atmospheric pressure while maintaining high transmittance.  
Ultra-high-molecular-weight polyethylene (UHMWPE) is widely used for such windows (e.g., ACT~\cite{Swetz2011ACT}, ABS~\cite{essinger2011probing}, MUSIC~\cite{golwala2012music}, QUIET~\cite{bischoff2013quiet}, CLASS~\cite{essinger2014class}, QUBIC~\cite{masi2022qubic}) thanks to its high tensile
strength
and low absorption (loss tangent $\tan\delta = 2 \times 10^{-4}$~\cite{essinger2011probing}) in the sub-THz band.  
We therefore adopt UHMWPE for the GLTCAM vacuum window.

To achieve high transmittance, optical reflection losses at the air--window and window--vacuum interfaces must be suppressed. Reflections reduce optical efficiency, introduce stray-light systematics, degrade polarization, and distort spectra via standing waves. Reflectance increases with refractive index $n$; for UHMWPE, although the index is relatively low ($n\approx 1.52$), the two interfaces still cause an average reflection loss of about 8\%. Therefore, broadband anti-reflection 
coatings (ARCs)
are essential.

Two main AR approaches are used in the sub-THz band: subwavelength 
structure (SWS) ARCs
and dielectric
multilayer ARCs
~\cite{abitbol2017cmb-s4,goldsmith1998quasioptical}.  
SWS ARCs
reduce the refractive-index mismatch at interfaces by gradually varying the effective index, using periodic microscopic grooves or pyramids much smaller than the wavelength.  
For example, CNC-machined SWSs on high-density polyethylene (HDPE) have achieved a 53\% fractional bandwidth~\cite{tapia2018systematic}, and 
SWS ARCs
with fractional bandwidths of 25--77\% on alumina and silicon have been fabricated using dicing saw machining
or deep reactive ion etching
~\cite{datta2013large,nitta2014anti,hasebe2021fabrication}.
The bandwidth values quoted here are provided as reference values.
Note that the transmission or reflection criteria used to define the band edges are not always explicitly specified and may differ between studies.
However, achieving bandwidths exceeding 100\% 
with SWS ARCs
remains  challenging
because
(1) higher center frequencies require finer features, and  
(2) suppressing diffraction-induced reflections at high frequencies demands smaller periods and thus larger aspect 
ratios.
Nevertheless, a fractional bandwidth of 116\% with 
transmission exceeding 97\% has been demonstrated on sapphire with an SWS ARC fabricated by laser ablation~\cite{Takaku2020}.

Multilayer
ARCs, in contrast, stack dielectric layers with chosen refractive indices and thicknesses to cause destructive interference of reflections.  
A common design uses quarter-wave layers ($nd=\lambda/4$, with thickness $d$ and target wavelength $\lambda$) arranged in a geometric progression of $n$ from vacuum to substrate.  
Porous polytetrafluoroethylene (PTFE) sheets ($n \sim 1.2$) have been widely used for sub-THz ARCs on UHMWPE and HDPE~\cite{benford2003,kooi2008advanced,zhang2009new,tran2009optical,arnold2010polarbear,lange2011highly,essinger2011probing,golwala2012music,bischoff2013quiet,nadolski2020broadband}.  
For a single-layer porous PTFE ARC on a UHMWPE substrate, the achievable fractional bandwidth is 74\% and 48\% when the band edges are defined by reflection-loss levels of 5\% and 2\%, respectively.
However, no porous PTFE materials provide the intermediate refractive indices and thicknesses needed to reach 130--710~GHz (138\%).

One promising solution is a dielectric multilayer synthesis approach~\cite{pisano2018}, which creates layers with intermediate effective refractive indices by stacking materials of different $n$ and of smaller thickness than required for a single layer.
In this study, we apply this approach using thin porous PTFE sheets as low-index materials 
and polyethylene glue layers as high-index materials, to synthesize the dielectric layers required for broadband ARCs on UHMWPE vacuum windows, thereby realizing the refractive indices and thicknesses unavailable from single commercial materials.

This paper presents an implementation of dielectric multilayer synthesis that achieves broadband anti-reflection performance and is suitable for large-aperture sub-THz vacuum windows.
Section~\ref{sec:survey} describes our material survey, Section~\ref{sec:AR_design} our design approach, Section~\ref{sec:fab} the fabrication process, and Section~\ref{sec:meas} the measured transmittance.  
Finally, Section~\ref{sec:con} summarizes the results and discusses future work.

\section{MATERIAL SURVEY AND CHARACTERIZATION}
\label{sec:survey}

We surveyed commercially available porous PTFE sheets with various porosities, pore sizes, and thicknesses to identify suitable materials for multilayer
ARCs on UHMWPE vacuum windows used in the GLTCAM cryostat.  
We require the GLTCAM vacuum window to provide
an average reflection loss below 2\% and a maximum 
reflection
loss below 4\% across 130--710~GHz.  
These values are chosen such that the reflection loss is reduced to a level comparable to the average (2\%) and maximum (4\%) absorption loss of a 9-mm-thick UHMWPE vacuum-window substrate over 130--710~GHz, where the substrate thickness is determined by mechanical requirements.

Assuming $n=1.52$ for UHMWPE~\cite{bischoff2013quiet,wilson2008aztec} and a center frequency of 420~GHz, 
designs based on a geometric progression profile of refractive indices require at least a 4-layer ARC to achieve the target in-band reflectance over the required bandwidth.
In this case, the target pairs of refractive index $n$ and thickness $d$ for the four layers from vacuum to substrate are ($n$, $d$) = (1.09, 164~{\textmu}m), (1.18, 151~{\textmu}m), (1.29, 139~{\textmu}m), and (1.40, 128~{\textmu}m).

However, obtaining materials with such precise combinations is challenging, and our literature survey revealed no low-loss dielectrics with $n=1.09$ or $n=1.40$ at thicknesses of $\sim 100$~{\textmu}m.  
We therefore conducted a survey of commercially available porous PTFE sheets to identify products whose optical properties are close to the target ranges.

We measured the refractive index $n$, extinction coefficient $k$, and thickness $d$ of selected porous PTFE products, Zitex G~\cite{zitex}, Poreflon FP/WP~\cite{poreflon}, and ADVANTEC T~\cite{advantec}, all having nominal thicknesses on the order of 100~{\textmu}m.  
Table~\ref{tab:materials} summarizes the selected products and their measured properties.  
The extinction coefficients of all surveyed materials were smaller than $1\times10^{-3}$ over the measured frequency range and are therefore not listed individually in the table.
Because all samples had pore sizes below 6~{\textmu}m, smaller than 1/50 of the wavelengths in our frequency range, scattering losses~\cite{benford2003} are expected to be negligible.  
Optical characterization was carried out using a 
terahertz time-domain spectroscopy (THz-TDS; ADVANTEST TAS7500TS), comprising a THz source module (TAS1100) and detector module (TAS1230) with a frequency resolution of 3.8~GHz across 0.1--4~THz.  
Optics with $F=3.75$ were placed in a sealed chamber, and samples were positioned at the focus. 
Holders with 26-mm apertures (approximately twice the 100~GHz beam waist diameter of $2w_{0}=14.3$~mm) were used for both sample and background measurements.  
Humidity was kept below 3\% RH by continuously supplying dry air to minimize errors from water vapor absorption.
To suppress fluctuations on $\sim 100$~s timescales, sample and background measurements were alternated every 30~s (1024 THz pulses each), and 50 interleaved pairs were acquired to reduce statistical noise.  

The measured transmittances were highly reproducible, with variations below 0.5\% over 100--800~GHz (see Uno et al.~\cite{uno2020demonstration} for details).  
The refractive index $n$ and extinction coefficient $k$ were then derived from the measured thickness, transmittance, and phase shift using the standard equations in Hangyo et al.~\cite{hangyo2002spectroscopy}.

We identified porous PTFE materials with refractive indices as low as $n \approx 1.08$ and thicknesses below 100~{\textmu}m, which to our knowledge have not been previously reported~\cite{benford2003,lange2011highly}.  
However, no materials with $n \approx 1.40$ were found.

The refractive index $n_{\textrm{LL}}$ of porous PTFE can be estimated from its porosity $p$ via the Lorentz--Lorenz (Clausius--Mossotti) relation~\cite{choy2015effective}:
\begin{equation}
\label{eq:LL}
    \frac{n_{\textrm{LL}}^{2}-1}{n_{\textrm{LL}}^{2}+2}
    = (1-p)\frac{n_{1}^{2}-1}{n_{1}^{2}+2},
\end{equation}
where $n_{1}$ is the refractive index of bulk PTFE.  
A comparison between the measured refractive indices and those estimated using Eq.~\ref{eq:LL} showed agreement 
within 6\% for all the samples, 
indicating that porosity is a reliable predictor of $n$ for material selection.

{\footnotesize
\begin{center}
\begin{threeparttable}[ht]
 \caption{Physical and Optical Properties of the Surveyed Porous PTFE Materials}
 \centering
  \begin{tabular}{cccccccc}
   \hline
   Product & Model & Thickness & Pore size & Porosity & $n_{\textrm{LL}}$\tnote{a} & $n$\tnote{b} \\
   && ({\textmu}m) & ({\textmu}m) & (\%) & & \\
   \hline
   Zitex & G104 & 102 & 5 -- 6 & 55 & 1.18 & 1.25 \\
   & G106 & 153 & 4 -- 5 & 50 & 1.20 & 1.25 \\
   & G108 & 220 & 3 -- 4 & 45 & 1.22 & 1.23 \\
   & G115 & 422 & 1 -- 2 & 40 & 1.25 & 1.23 \\
   \hline
   Poreflon & FP-010-60 & 72 & 0.1 & 55 -- 70 & 1.12 -- 1.18 & 1.23 \\
   & FP-030-200 & 206 & 0.3 & 30 -- 40 & 1.25 -- 1.29 & 1.27 \\
   & FP-050-300 & 302 & 0.5 & 70 & 1.12 & 1.17 \\
   & FP-100-100 & 76 & 1 & 60 -- 75 & 1.10 -- 1.16 & 1.13 \\
   & WP-020-80 & 77 & 0.2 & 60 -- 75 & 1.10 -- 1.16 & 1.12 \\
   & WP-045-80 & 72 & 0.45 & 60 -- 75 & 1.10 -- 1.16 & 1.10 \\
   & WP-100-100 & 80 & 1 & 65 -- 80 & 1.08 -- 1.14 & 1.08 \\
   \hline
   ADVANTEC & T010A & 84 & 0.1 & 68 & 1.13 & 1.19 \\
   & T020A & 86 & 0.2 & 74 & 1.10 & 1.14 \\
   & T050A & 91 & 0.5 & 78 & 1.09 & 1.13 \\
   & T080A & 68 & 0.8 & 76 & 1.09 & 1.11 \\
   & T100A & 70 & 1 & 79 & 1.08 & 1.10 \\
   & T300A & 74 & 3 & 83 & 1.07 & 1.08 \\
   \hline
  \end{tabular}
  \label{tab:materials}
\begin{tablenotes}
    \item[a] $n_{\textrm{LL}}$ is the refractive index estimated using Eq.~\ref{eq:LL}, assuming bulk PTFE has $n=1.43$~\cite{lamb1996}.
    \item[b] $n$ and $k$ are the measured refractive index and extinction coefficient at 300~GHz, respectively.
\end{tablenotes}
\end{threeparttable}
\end{center}
}

\section{ARC DESIGN AND OPTIMIZATION}
\label{sec:AR_design}
In Section~\ref{sec:survey}, we found porous PTFE sheets with low $n$ but not the exact $(n,d)$ combinations required for the multilayer design.  
To address this, we adopt a multilayer synthesis approach~\cite{pisano2018}, in which high- and low-index materials are laminated to produce intermediate effective refractive indices.
This approach requires porous PTFEs that are thinner and have lower $n$ than conventionally used, as obtained in Section~\ref{sec:survey}, and enables a broader range of layer combinations than previously available.

For the high-index component, the selection criteria are low dielectric loss, a refractive index close to that of the substrate, and ready availability in various thicknesses.
Low-density polyethylene (LDPE) and high-density polyethylene (HDPE), with $n=1.52$ and $\tan\delta \approx 3\times10^{-4}$~\cite{lamb1996}, satisfy these criteria and are readily available in a variety of thicknesses.
Although LDPE has conventionally been used only as an adhesive, with its thickness minimized to avoid design complexity and changes during thermal bonding~\cite{tran2009optical, lange2011highly, kooi2008advanced, essinger2011probing, bischoff2013quiet, golwala2012music, essinger2014class, hargrave2010anti, nadolski2020broadband}, here we propose to employ LDPE/HDPE glue layers as high-index dielectrics.
With appropriate thickness selection and bonding control (see Section~\ref{subsec:thick_cont}), these layers can serve a dual role as both adhesive and high-index dielectric, enabling multilayer synthesis without introducing additional materials.

This design approach provides robustness against both manufacturing variations and thickness changes during the ARC fabrication process, since the effective index can be fine-tuned by adjusting the glue thickness. 
Therefore, optimization remains feasible even if the porous PTFE deviates from its nominal design value.

We optimized the ARC on a UHMWPE ($n=1.52$) substrate by numerical calculations as follows.
The substrate thickness was set to 2~mm so that interference fringes from multiple reflections between its interfaces would be resolvable by our THz-TDS (3.8~GHz resolution).
The design requirement was set to
a reflection loss $<2\%$ (average) and $<4\%$ (maximum) over 130--710~GHz. 

Up to five porous PTFE layers ($L_\mathrm{p1}$--$L_\mathrm{p5}$) were placed on each side of the substrate, with glue layers ($L_\mathrm{g1}$--$L_\mathrm{g4}$) inserted between them, and the same stack was applied on both sides.
The refractive indices used in the optimization were taken from the measured material properties summarized in Table~\ref{tab:materials}, and the extinction coefficients were assumed to be negligible.
For pPTFE films, whose thickness varies between production lots, the layer thicknesses were fixed to those of the pPTFE stock prepared for ARC fabrication.
The porous PTFE layers included PTFE ($n = 1.43$), whose $n$ is closest to that of the substrate-side layer.
The PTFE thicknesses were selected from commercially available films with nominal thicknesses of 80, 100, and 130~{\textmu}m.
Glue layers were assigned $n=1.52$ (LDPE / HDPE), 
and their thicknesses were discretely varied from 5--50~{\textmu}m in 5~{\textmu}m steps in the optimization, reflecting the practical thickness increments of commercially available products.
The in-band transmission was calculated using the transfer-matrix method~
\cite{yeh1988optics}
implemented in \texttt{ctmm}~\cite{ctmm}, and the designs meeting the requirements were selected.

\begin{figure}[ht]
    \centering
    \begin{subfigure}{0.47\linewidth}
        \includegraphics[width=\linewidth]{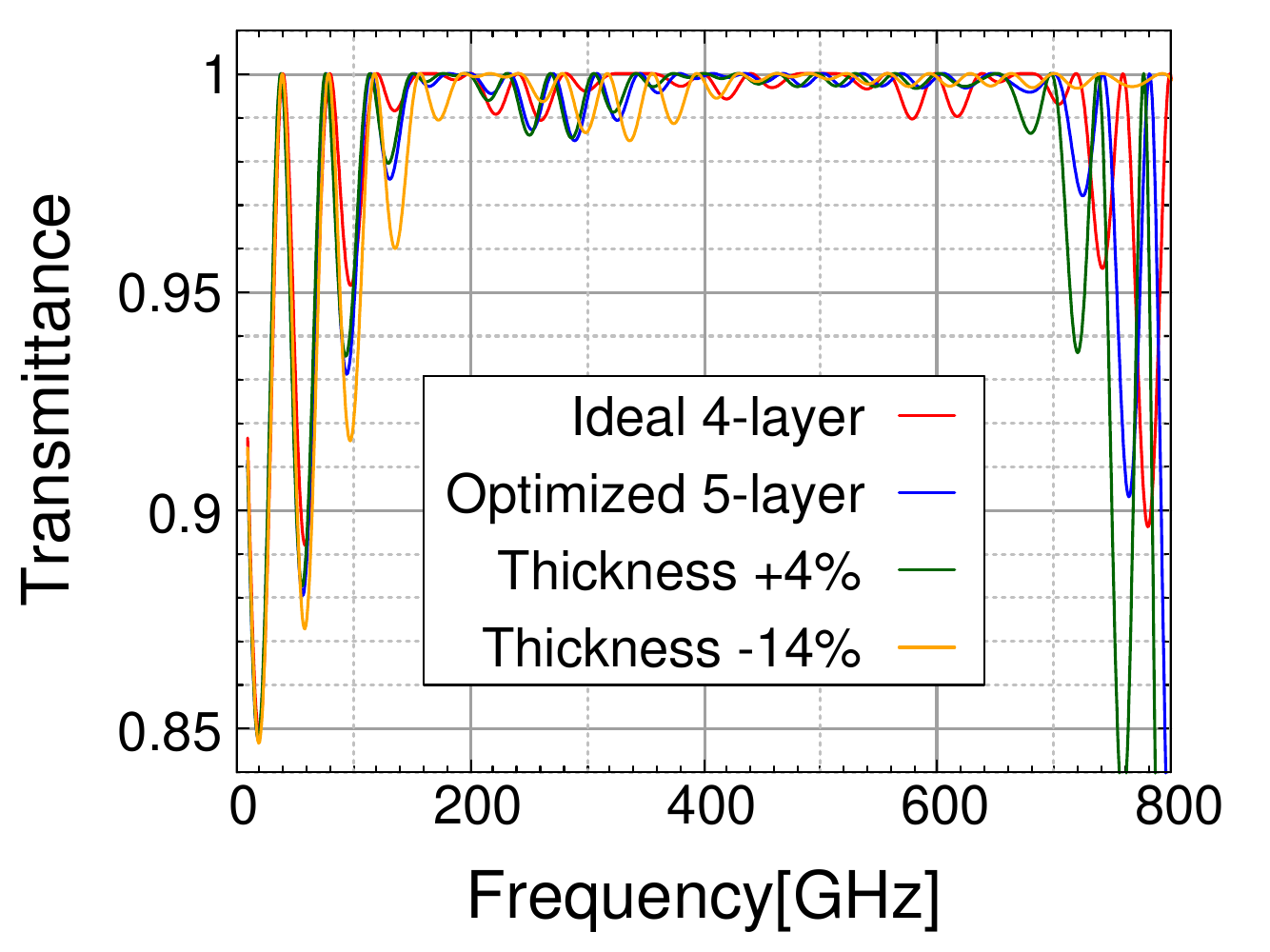}
        \caption{}\label{fig:vs_ideal}
    \end{subfigure}
    \begin{subfigure}{0.47\linewidth}
        \includegraphics[width=\linewidth]{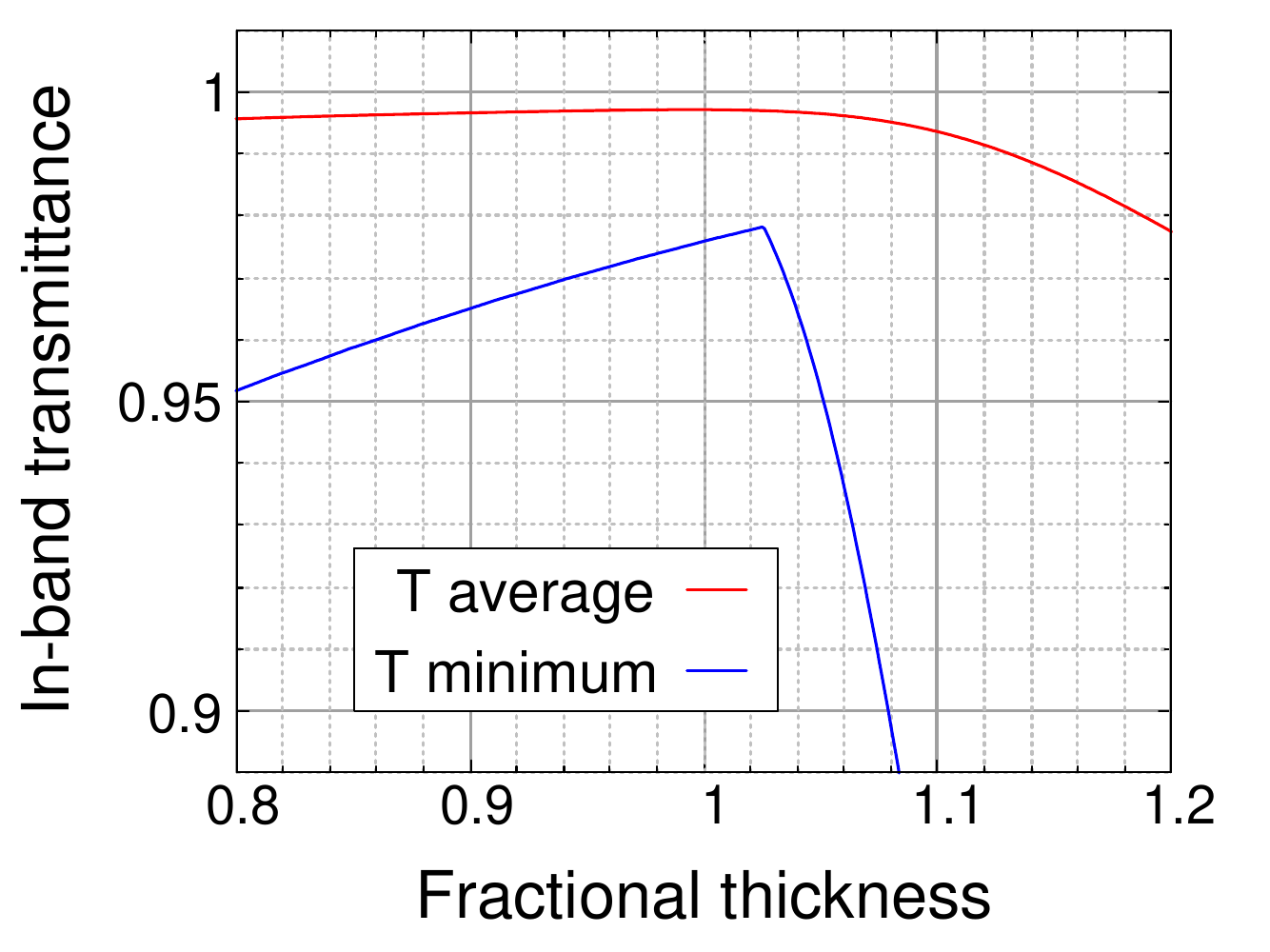}
        \caption{}\label{fig:tolerance}
    \end{subfigure}
    \caption{(a) Calculated transmittance spectra of the (red) 
    4-layer ARC based on a geometric progression of refractive indices
    and the (blue) optimized 5-layer ARC (Table~\ref{tab:ARdesign}) on a 2-mm-thick UHMWPE substrate. 
    Spectra with uniform thickness deviations of +4\% and $-$14\% across all layers (green and yellow) demonstrate tolerance to thickness variation. 
    These calculations assume no material absorption ($\tan\delta = 0$). 
    (b) Calculated (red) average and (blue) minimum in-band transmittances (130--710~GHz) as a function of uniform thickness variation from the design values (Table~\ref{tab:ARdesign}).}
    \label{fig:Tdesign}
\end{figure}

Among the candidates, we obtained an optimized design consisting of five porous PTFE layers (Table~\ref{tab:ARdesign}) 
that achieved an average in-band reflection loss of 0.3\% within 130--710~GHz.  
It also met the $<4\%$ maximum-loss criterion over 103--752~GHz (152\% fractional bandwidth), with an average loss of 0.4\%, comparable to
a 4-layer ARC with a geometric progression of refractive indices
(Fig.~\ref{fig:Tdesign}\subref{fig:vs_ideal}).

The optimal design can be regarded as an approximation to
a 5-layer ARC design with a geometric progression of refractive indices,
from which the layer with $n=1.32$ is omitted,
($n$, $d$) = (1.07, 166~{\textmu}m), (1.15, 155~{\textmu}m), (1.23, 145~{\textmu}m),
(1.32, 135~{\textmu}m), and (1.42, 126~{\textmu}m), 
yielding a maximum reflection loss of 1.6\% and an average of 0.4\% over 130--710~GHz.

\begin{center}
\begin{threeparttable}[ht]
 \caption{Optimized 5-Layer ARC Design for a 2-mm-Thick UHMWPE Substrate}
 \centering
  \begin{tabular}{ccccc}
   \hline
   Sheet layer& Glue layer& Material & Thickness & $n$ \\
   ($L_{\textrm{p}}$) No.& ($L_{\textrm{g}}$) No.&& ({\textmu}m) & \\
   \hline
   1 && ADVANTEC T300A\tnote{a} & 88 & 1.08 \\
   & 1 & HDPE & 5 & 1.52 \\
   2 && ADVANTEC T300A & 88 & 1.08 \\
   & 2 & HDPE & 20 & 1.52 \\
   3 && ADVANTEC T020A & 99 & 1.14 \\
   & 3 & HDPE & 30 & 1.52 \\
   4 && Poreflon FP-010-60 & 79 & 1.23 \\
   & 4 & HDPE & 15 & 1.52 \\
   5 && PTFE\tnote{a} & 100 & 1.43 \\
   \hline
   & - & LDPE\tnote{a} & 50 & 1.52 \\
   Window && UHMWPE\tnote{a} & 2000 & 1.52 \\
   \hline
  \end{tabular}
  \label{tab:ARdesign}
\begin{tablenotes}
 \item[a] The optimization was performed with these parameters fixed.
\end{tablenotes}
\end{threeparttable}
\end{center}

Finally, Fig.~\ref{fig:Tdesign}\subref{fig:tolerance} shows the dependence of average and minimum in-band transmittance on uniform thickness variation across all layers, representing possible errors from lot-to-lot variations or mechanical compression during bonding.
The required performance is maintained for variations within $-14\%$ to $+4\%$, with the minimum transmittance being more sensitive.

Based on this optimized design, we proceeded to fabricate 5-layer ARC samples, as described in Section~\ref{sec:fab}.

\section{FABRICATION}
\label{sec:fab}

Before fabricating samples based on the 5-layer ARC design, we assessed feasibility and identified two potential issues:
\begin{itemize}
    \item Compression during bonding, which can reduce the thickness of the AR materials
    \item Heating, which can cause dimensional changes and warpage in the substrate
\end{itemize}
Such thickness reduction would degrade performance, while substrate deformation could create voids and weaken bonding. 
We therefore developed a fabrication process to minimize thickness reduction (Section~\ref{subsec:thick_cont}) and thermal deformation (Section~\ref{subsec:mitig}) before fabricating the samples (Section~\ref{subsec:sample_fab}). 
Each sample measured 30~mm $\times$ 30~mm, suitable for THz-TDS evaluation.

\subsection{Layer thickness control}
\label{subsec:thick_cont}

Bonding requires pressure to ensure adhesion and suppress air bubbles, but this inevitably compresses the AR materials. 
Hot pressing, a common method~\cite{dumoulin2011results, bischoff2013quiet, essinger2014class}, can reduce thickness by up to 25\%~\cite{pisano2018}, exceeding our 14\% tolerance (Section~\ref{sec:AR_design}). 
We therefore tested low-pressure bonding. 
A 160~g load on 30~mm $\times$ 30~mm samples ($\approx 2$~kPa~\cite{essinger2011probing}) was applied using HDPE films as glue between porous PTFE sheets. 
Samples were heated at 150~{\textcelsius} for 1~hour, then slowly cooled to room temperature over 12~hours. 
The total thickness decreased by only 3\%, within tolerance. 
No air bubbles or peeling appeared, even after 10 rapid thermal cycles between 77~K and 296~K by repeated immersion in liquid nitrogen.
Thus, $\approx 2$~kPa bonding achieves adequate adhesion while limiting thickness reduction.

\subsection{Thermal deformation mitigation}
\label{subsec:mitig}

Annealing is effective in suppressing thermal deformation~\cite{rosato2012injection}. 
We set the bonding temperature to 133~{\textcelsius}, below the melting point of UHMWPE ($\sim 135$~{\textcelsius}) but above that of LDPE ($\sim 110$~{\textcelsius}).
However, this temperature exceeds the heat-deformation temperatures of UHMWPE ($\sim 96$~{\textcelsius}) and porous PTFE ($\sim 120$~{\textcelsius}), risking deformation from the release of residual manufacturing stresses. 
To mitigate this, we annealed UHMWPE and porous PTFE using the same temperature cycle as the bonding process, with slow cooling to suppress the formation of new residual stresses.
As a result, UHMWPE shrank by up to 3\% in-plane and expanded by up to 6\% in thickness, while porous PTFE shrank by up to 7\% in-plane and showed thickness changes within the 1\% measurement accuracy.
No further deformation was observed after re-annealing.

During bonding, through-thickness and in-plane temperature gradients within UHMWPE caused warpage upon cooling and uneven melting that led to non-uniform bonding. 
To suppress these effects, samples were sandwiched between metal plates with high thermal conductivity and similar heat capacities, and a programmable forced-convection oven was used to maintain top-bottom temperature differences below 1~{\textcelsius}.
In addition, slow cooling from 133~{\textcelsius} to 80~{\textcelsius} (within its deformation temperature range) over 8~hours reduced warpage to $<30$~{\textmu}m over 50~mm for 5-mm-thick UHMWPE, meeting our requirement.

\subsection{Fabrication of ARC-coated samples}
\label{subsec:sample_fab}

\begin{figure}[ht]
\centering\includegraphics[width=7cm]{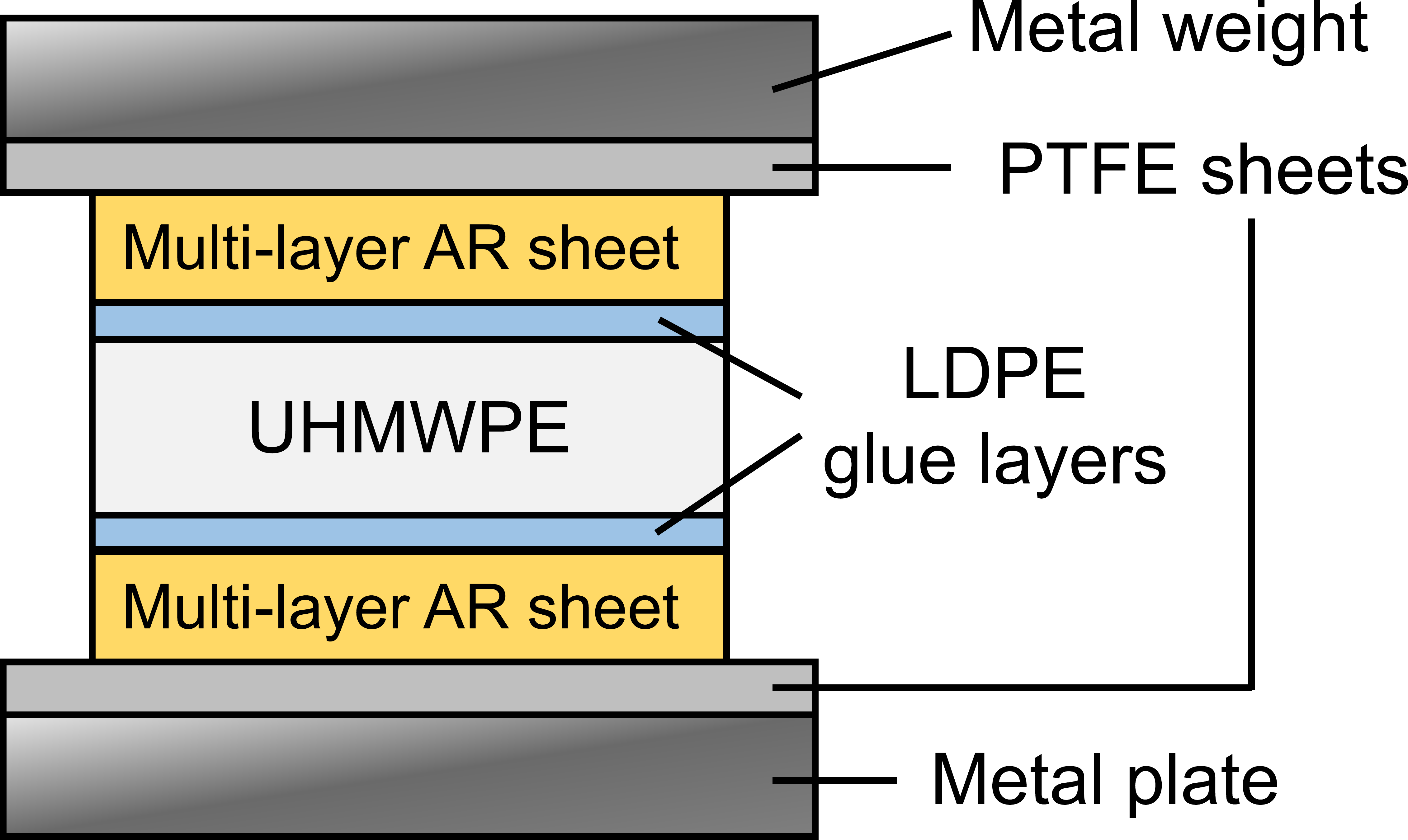}
\caption{Schematic illustration of the bonding method used to apply the ARC to a vacuum window.}
\label{fig:heating_config}
\end{figure}

We fabricated AR-coated samples for THz-TDS evaluation following the above procedures and the design parameters in Table~\ref{tab:ARdesign}. 
To evaluate changes in thickness and transmittance caused by bonding, the process was divided into two steps. 
First, we fabricated two multilayer AR sheets for both sides. 
To prevent re-melting of the glue during bonding to the substrate, which could change glue thickness, we used HDPE (with a higher melting point than LDPE but lower than UHMWPE) as the glue within the AR sheets. 
The thickness of the fabricated 5-layer AR sheets decreased by 3\%, consistent with the results in Section \ref{subsec:thick_cont}. 
Next, we bonded these AR sheets to both sides of a 2-mm-thick UHMWPE substrate prepared by machining both sides of an annealed 5-mm-thick slab to ensure uniform thickness, using the oven setup shown in Fig.~\ref{fig:heating_config}. 
PTFE sheets with low friction were placed between the metal plates and AR sheets to allow smooth sliding and prevent interface dragging from thermal expansion mismatch.

We also applied this method to a 6-mm-thick, 140-mm-diameter UHMWPE vacuum window and successfully bonded AR coatings to both sides without warpage or adhesion problems. 
This 
provides more confidence that our method can be extended
to larger-diameter vacuum windows, including the 300-mm window required for the GLTCAM cryostat.

\begin{figure}[ht]
\centering\includegraphics[width=7cm]{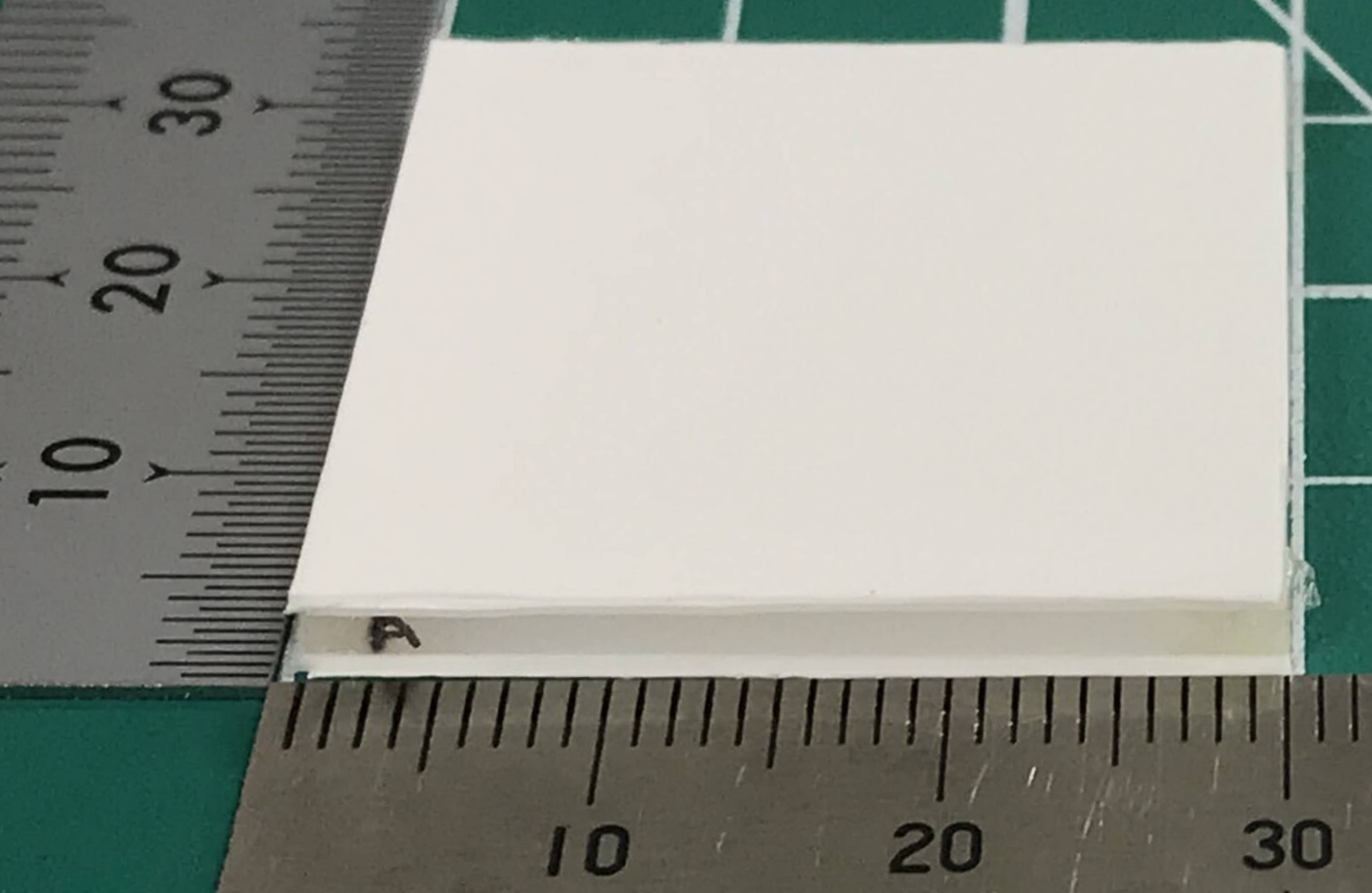}
\caption{Photograph of the fabricated AR-coated UHMWPE slab using the method described in this section.}
\label{fig:fabricated_sample}
\end{figure}

The fabricated AR-coated sample is shown in Fig.~\ref{fig:fabricated_sample}.

\section{TRANSMITTANCE MEASUREMENTS AND DISCUSSION}
\label{sec:meas}

\begin{figure}[ht]
\centering\includegraphics[width=7cm]{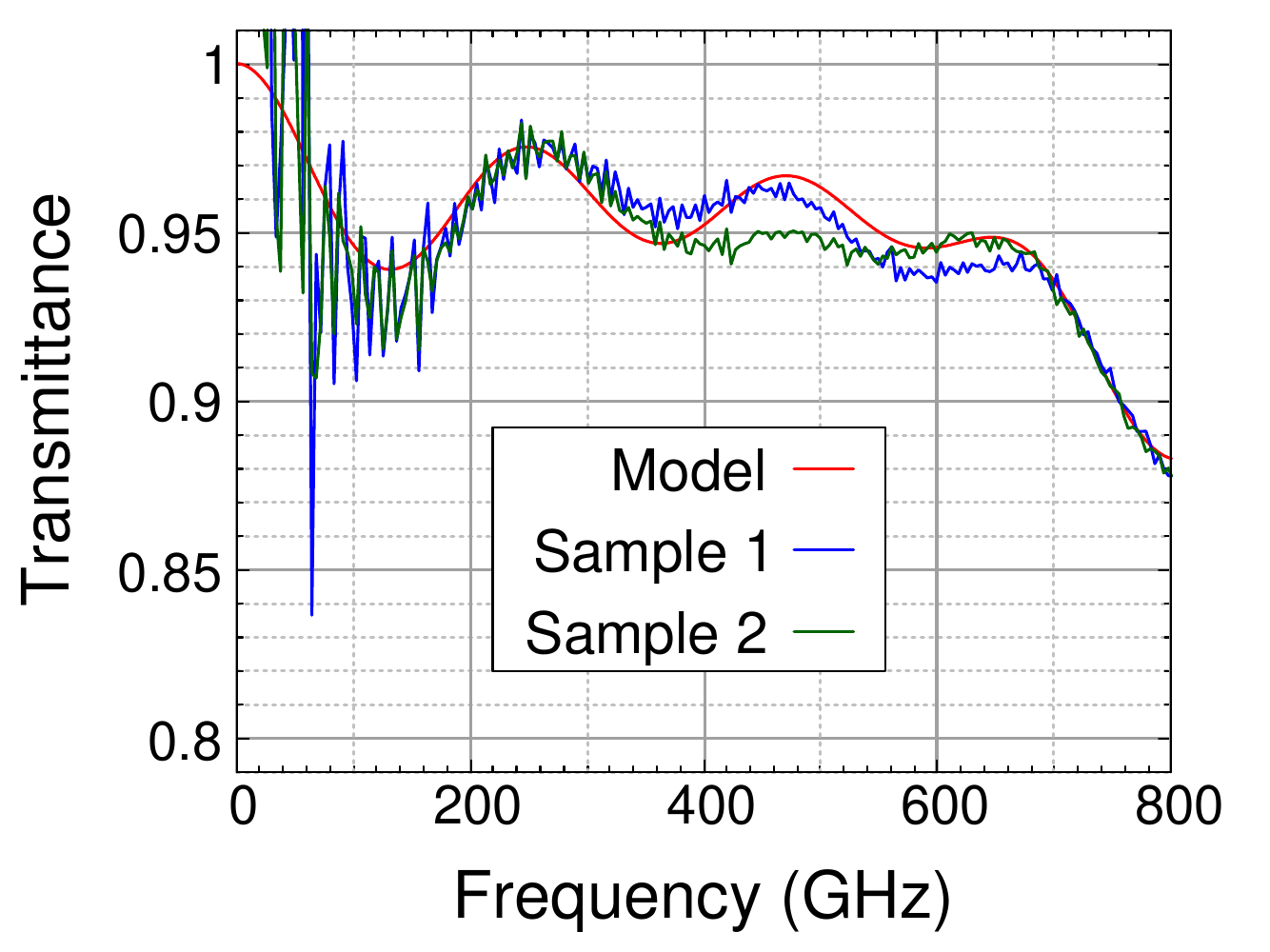}
\caption{Measured transmittance spectra (blue and green) of two fabricated 5-layer AR sheets, compared with the model transmittance spectrum (red) calculated from the design values (Table~\ref{tab:ARdesign}). 
The model does not include material absorption, which is negligible.}
\label{fig:meas_sheets}
\end{figure}

We evaluated the transmittance of two fabricated 5-layer AR sheets at 296~K using THz-TDS. 
Fig.~\ref{fig:meas_sheets} shows the measured spectra together with the lossless model curve based on the design in Table~\ref{tab:ARdesign}. 
The measured transmittances agree with the model within $\sim 1$\%, demonstrating the validity of the design and fabrication processes as well as the reproducibility of the samples. 
The results also indicate that in-band (130--710~GHz) absorption of the AR sheets is negligible. 
As expected, the 3\% compression of the AR sheets had a negligible effect on their optical performance.

\begin{figure}[ht]
\centering\includegraphics[width=7cm]{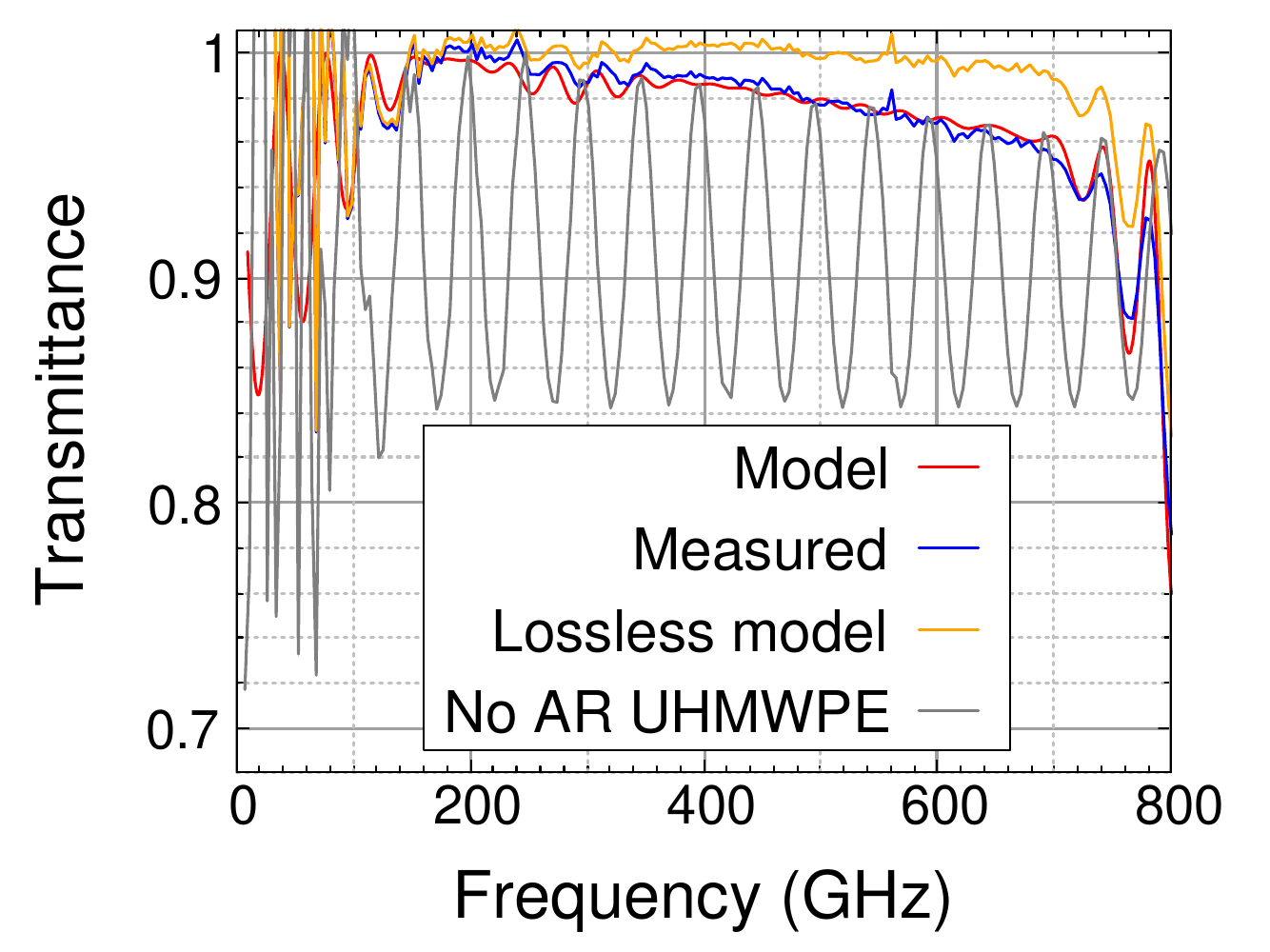}
\caption{
Transmittance spectrum of a 5-layer AR-coated 2-mm-thick UHMWPE slab. 
Blue: measured transmittance by THz-TDS. 
Red: model spectrum based on the design values (Table~\ref{tab:ARdesign}), including UHMWPE absorption. 
Yellow: lossless spectrum derived by subtracting the UHMWPE absorption model from the measured spectrum. 
Gray: measured spectrum of the bare UHMWPE slab before applying the ARC.}
\label{fig:meas_ARcoated}
\end{figure}

We further measured the transmittance of a 2-mm-thick UHMWPE slab coated with these AR sheets at 296~K. 
Fig.~\ref{fig:meas_ARcoated} shows the measured spectrum, together with the model prediction. 
In the model, only absorption in the UHMWPE slab is considered, as absorption in the AR sheets is negligible. 
The UHMWPE absorption model was derived by fitting a third-order polynomial to the envelope of the transmittance spectrum measured before applying the ARC. 
The measured in-band characteristics of the AR-coated UHMWPE closely match the model, while fluctuations below 100~GHz are attributed to the low signal-to-noise ratio and are outside the nominal observation band.

Finally, we estimated the reflection loss of the AR-coated UHMWPE slab using the lossless model in Fig.~\ref{fig:meas_ARcoated}, which ignores UHMWPE absorption. 
The analysis yielded an average in-band reflection loss of 0.2\% and a maximum loss of 3.2\%, satisfying the design requirements of $<2$\% and $<4$\%, respectively. 
We note that the transmittance appears to exceed 100\% in the 200--500~GHz range; however, this is within the systematic error margin, considering the $\sim 0.5$\% measurement uncertainty of THz-TDS. 
These results demonstrate that the developed multilayer ARC provides 
broadband anti-reflection performance suitable for vacuum windows, validating both the proposed design concept and the fabrication method.

\section{CONCLUSIONS}
\label{sec:con}

Next-generation sub-THz astronomical instruments require vacuum windows with high transmission over a broad band of frequencies. In this study we presented a multilayer ARC technique for a UHMWPE vacuum window that satisfies the requirement of GLTCAM.

Through
THz-TDS
measurements, we identified thin, low-index porous PTFE materials ($n \sim 1.1$, $d<100$~{\textmu}m), which had not been reported previously.  
However, no materials with $n \sim\ 1.3$ and $d>100$~{\textmu}m, required for broadband 
multilayer
ARC designs, were available.  
To overcome this, we
used
a design method based on dielectric multilayer synthesis that combines the newly identified porous PTFEs with glue layers as high-index components.  
Numerical
calculations yielded an optimized 5-layer design achieving an average reflection loss of 0.3\% and a maximum reflection loss of 1.6\% over 130--710~GHz, and meeting the $<4$\% maximum-loss criterion over 103--752~GHz (152\% fractional bandwidth) with a 0.4\% average loss.

We then established a fabrication process and demonstrated AR-coated UHMWPE windows whose measured reflection losses were 0.2\% (average) and 3.2\% (maximum) over 130--710~GHz, fully satisfying the design requirements.  
The thin porous PTFEs identified in this study 
enables
flexible dielectric layer synthesis for even broader or shifted spectral bands, paving the way for next-generation 
broadband
sub-THz instrumentation.  
We are currently extending this technique to large apertures and curved surfaces for application to the GLTCAM vacuum window and dielectric lenses.

\section*{Funding.}
Japan Society for the Promotion of Science KAKENHI (JP17H02872, JP19K14754, JP20H01937, JP23H00121, JP23K25879, JP23K25905, JP24K22911);
NAOJ Research Coordination Committee, NINS (NAOJ-RCC-2201-0101);
Murata Science and Education Foundation; Nakajima Foundation; 
Sumitomo Foundation (Basic Science Research Grant No. 2200541).

\section*{Acknowledgments.}
This study was carried in cooperation with the Advanced Technology Center of the National Astronomical Observatory of Japan (NAOJ).
S.~U. was supported by RIKEN Special Postdoctoral Researcher Program.
S.~I. was supported by FoPM, WINGS Program, the University of Tokyo.
K.~W. was supported by JST SPRING, Japan Grant Number JPMJSP2104.
T.~T. was supported by MEXT Leading Initiative for Excellent Young Researchers Grant Number JPMXS0320200188.

\section*{Disclosures.}
The authors declare no conflicts of interest.

\section*{Data availability.}
Data underlying the results presented in this paper are not publicly available at this time but may be obtained from the authors upon reasonable request.


\bibliography{references}

\end{document}